\documentclass[doublecol]{epl2}

\usepackage{epsfig,graphicx}

\usepackage{amssymb}
\usepackage{amsmath}
\usepackage{dcolumn}
\usepackage{bm}

\title{The effect of uniaxial pressure on the magnetic anisotropy of the Mn$_{12}$-Ac single-molecule magnet}
\shorttitle{The effect of uniaxial pressure on the magnetic anisotropy of Mn$_{12}$-Ac}

\author{James H. Atkinson\inst{1,2} \and K. Park\inst{3} \and C. C. Beedle\inst{4} \and D. N.
Hendrickson\inst{4} \and Y.
Myasoedov\inst{5} \and E. Zeldov\inst{4}  \and Jonathan R.
Friedman\inst{1,6} }
\shortauthor{J. H. Atkinson \etal}
\institute{
\inst{1} Department of Physics,
Amherst College, Amherst, Massachusetts 01002, USA\\
\inst{2} Department of Physics, University of Central Florida, Orlando, Florida 32816, USA\\
\inst{3} Department of Physics, Virginia Tech, Blacksburg VA 24061, USA\\
\inst{4}Department of Chemistry $\&$ Biochemistry, UCSD, La
Jolla, California 92093, USA\\
\inst{5}Department of Condensed
Matter Physics, Weizmann Institute of Science, Rehovot 76100,
Israel\\
\inst{6}Institute for Quantum Computing, University of Waterloo,
Waterloo, Ontario, Canada,
N2L 3G1}


\abstract{
\noindent We study the effect of uniaxial pressure on the
magnetic hysteresis loops of the single-molecule magnet
Mn$_{12}$-Ac.  We find that the application of pressure along
the easy axis increases the fields at which quantum tunneling
of magnetization occurs.  The observations are attributed to an
increase in the molecule's magnetic anisotropy constant $D$ of
0.142(1)$\%$/kbar.  The increase in $D$ produces a small, but measurable increase in the effective energy barrier for magnetization reversal.  Density-functional theory calculations also predict an increase in the barrier with applied pressure.}

\begin{document}
\maketitle
\section{Introduction}

\begin{figure}[tbh]
\centering
\includegraphics[width=80mm]{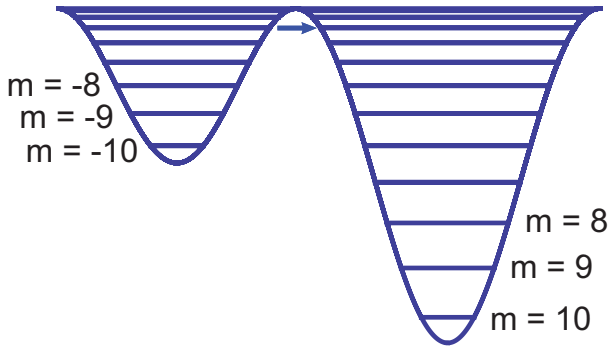} 
\caption{Double-well potential diagram illustrating the energy
levels for an S = 10 system.  A magnetic field tilts the
potential.  When the field brings levels in opposite wells into
alignment, the spins reverse via a process of resonant
tunneling.  At temperatures above $\sim$0.5 K, this process is
thermally assisted and tunneling occurs between levels near the
top of the barrier, as indicated by the arrow. \label{wells} }
\end{figure}

Single-molecule magnets (SMMs) are fascinating systems in
which each molecule behaves as a single rigid, high-spin
object. Most SMMs have a large magnetomolecular anisotropy that
impels the spin to point along a particular axis, the so-called easy axis.  Thus, the
spin's energy landscape can be described by a double-well
potential, with an energy barrier between ``up" and ``down"
orientations, as shown in Fig.~\ref{wells}. 

Much of SMMs' behavior can be understood from the simple
Hamiltonian~\cite{5}
\begin{equation}
\begin{aligned}
{\cal H} =  - DS_z ^2  - AS_z ^4  - g\mu _0\mu _B {\bf S} \cdot {\bf H} + {\cal H^\prime}
\label{ham}
\end{aligned}
\end{equation}
Here the first and second terms create the double-well
potential. The spin vector minimizes its energy by pointing
along or antiparallel to the easy z axis.  The constants $D$
and $A$ give the strength of the anisotropy, producing a
classical barrier of height $U=DS^{2} + AS^{4}$ at zero field.  
The third term is produced by the Zeeman interaction, with $g=1.96$.  When a field is
applied along the z axis, the potential is tilted, making, say,
the ``up" direction lower in energy than the ``down" direction. ${\cal H^\prime}$ contains terms that do not commute with $S_z$ and can therefore induce tunneling.

Because SMMs have a finite spin $S$ ($\sim$10), there are
$2S+1$ magnetic orientation states, $m=-S,-S+1,\ldots,S$,
associated with the spin (levels in Fig.~\ref{wells}).  As the
field along the easy axis is increased, levels in one well move
down while levels in the other well move up.  At certain fields~\cite{5},
\begin{equation}
\begin{aligned}
H_N=-H_{-N}  = \frac{{N}}{{g\mu _0 \mu _B }}\left[ {D +
A(m^2  + m'^2 )} \right] \label{nfield},
\end{aligned}
\end{equation}
where $N = -(m + m') = 0, \pm1, \pm2,\ldots$, levels $m$ and $m'$ in
opposite wells align, allowing the spin to tunnel between wells
and consequently producing a dramatic increase in the rate at
which the spins come to equilibrium~\cite{509}.  This effect,
which is manifested by the appearance of steps in the
hysteresis loops of most SMMs, has now been seen in hundreds of
molecules and some rare-earth ions~\cite{5}.

Mn$_{12}$ acetate (henceforth referred to as Mn$_{12}$) is the
most studied SMM. Crystals of Mn$_{12}$ typically contain two
species~\cite{397,393}: a fast relaxing minority species and a
slower relaxing majority species.  Here we exclusively study the latter, which has a spin $S = 10$,  a large
energy barrier of $\sim$70 K, and is well described by Eq.~\ref{ham} with $D=0.556$ K
and $A=1.13$ mK~\cite{852}.  Each magnetic molecule in a
crystal is isolated from its neighbors, resulting in a total
magnetic moment that is, to good approximation, an ensemble average of that of a single
Mn$_{12}$ molecule.  At temperatures above $\sim 0.5$ K, the relaxation between the two wells is thermally assisted, with tunneling taking place
from excited levels that are thermally populated.  In this regime, the magnetic relaxation rate $\Gamma$ is well described by an Arrhenius law~\cite{5}:

\begin{equation}
\Gamma=\omega_0 e^{-U_{eff}/T},
\label{arr}
\end{equation}

\noindent where $U_{eff}$ is the effective barrier, defined as the energy difference between the lowest level in the metastable well and the level $m$ where the preponderance of tunneling takes place, i.e. $U_{eff}=D\left(S^2-m^2\right)+A\left(S^4-m^4\right)+g\mu _0\mu _B H_z (-S-m)$;  $\omega_0$, the Arrhenius prefactor, is a constant of order 10$^7$ rad/s.

There have been numerous hydrostatic pressure experiments on
Mn$_{12}$ and related molecules~\cite{857, 865,858, 853, 866, 111, 854, 855, 856, 859}. Inelastic neutron scattering experiments have shown
a few-percent increase in $D$ when pressure of the order of a
dozen kbar is applied~\cite{865}. In contrast, Levchenko \emph{et al}.~\cite{858} reported that pressure increases the measured
magnetic relaxation rate, consistent with a decrease in $D$. In
addition, studies have found that pressure converts some of the
molecules between the fast and slower relaxing varieties,
although the precise behavior of this change remains unclear~\cite{858, 859, 111, 857}. In the present work, we apply
\emph{uniaxial pressure} along the easy axis of a single
crystal of Mn$_{12}$ and examine the energy-level structure
through measurements of the sample's hysteresis loops.  We find
that the pressure alters the field at which tunneling occurs,
an effect we interpret as due to an increase in $D$.  We also find a slight decrease in $\Gamma$ with pressure, which we attribute to the increase in $U_{eff}$ brought about by the increase in $D$.  Our
technique allows us to vary the pressure \emph{in situ} at low
temperature so that the sample is never under high pressure at
room temperature, minimizing any structural distortions and
deformations that may occur at such temperatures. 

\begin{figure}[bht]
\centering
\includegraphics[width=70mm]{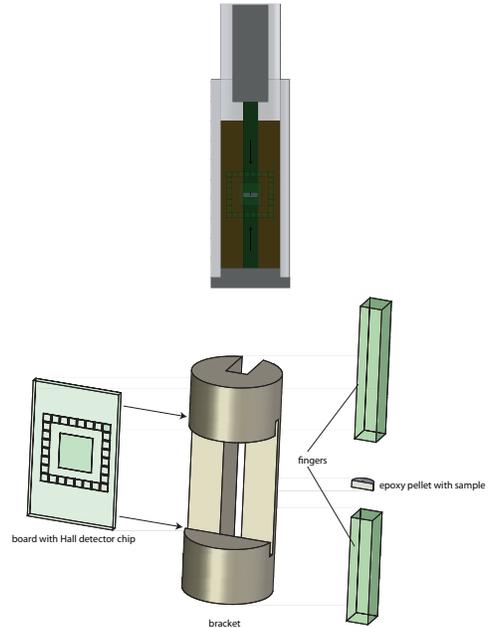} 
\caption{(top) Cross section schematic of our uniaxial-pressure cell. (bottom) Exploded view of the
cell core, showing Hall detector chip mounted on circuit board, the bracket, G10
fingers, and the epoxy pellet containing the embedded sample.
\label{apparatus} }
\end{figure}
\section{Experimental Setup}

Our uniaxial pressure apparatus is based on a design by Campos
\emph{et al}.~\cite{860}.  A diagram of the low-temperature
portion of the apparatus is shown in Fig.~\ref{apparatus}.
Pressure is applied through a pneumatic piston connected to a
nitrogen gas cylinder, and is transferred to the sample cell at
the end of the probe by non-magnetic stainless steel rods,
capped by a small aluminum plug. 
This plug and the bottom cap of the cell
press on two G10 phenolic ``fingers" that transmit the pressure
to a single crystal of Mn$_{12}$, which is encased in a small
(non-circular) cylindrical piece of epoxy (Stycast 1266), as
shown in Fig.~\ref{apparatus}.

The sample was prepared by placing it in freshly mixed epoxy
within a Teflon mold.  A 3-Tesla field was applied to align the
crystal's easy axis (crystallographic c axis) with the axis of
the cryostat's field.  Once cured, the epoxy was machined to
make the top and bottom horizontal faces flat and to create a
flat vertical face with the sample close to the surface,
allowing coupling to a Hall bar detector.

The bottom portion of Fig.~\ref{apparatus} illustrates the bracket that holds the
Hall sensor chip and its mounting board, the G10 fingers, and the epoxy
``pellet" containing the sample. When everything was in place,
one edge of the sample was aligned with one Hall sensor;
another sensor relatively far from the sample was used to
measure background signals. The signals from the two Hall bars
were subtracted using an analog circuit in order to cancel out
common background signals.  A regulated thermometer was placed
close to the epoxy-sample pellet for temperature measurement and
control.

\begin{figure}[thb]
\centering
\includegraphics[width=85mm]{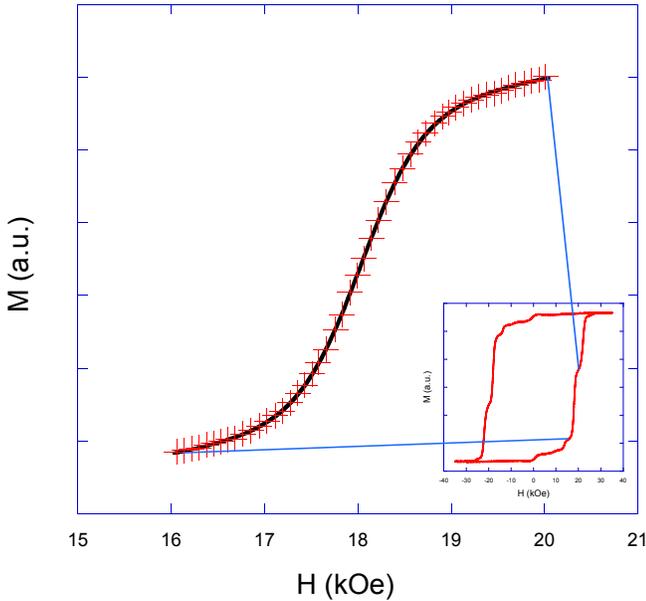} 
\caption{Detail of the N = 4 step with fit.  The data was taken
at 2.0 K and with no pressure applied.  The inset shows the
full hysteresis loop from which the detail was taken.
\label{inset} }
\end{figure}
\section{Results and Analysis}

\begin{figure}[hbt]
\centering
\includegraphics[width=85mm]{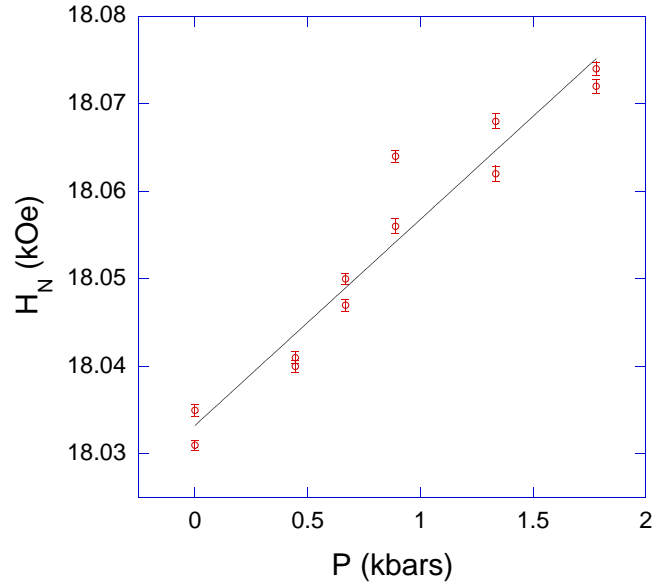} 
\caption{Step center field values as a function of pressure for
the N = 4 step at 2.0 K.  The line is the result of a linear
fit of the data. \label{hp2k} }
\end{figure}

We acquired magnetic hysteresis loops from the sample at
temperatures between 1.8 K and 2.2 K in 50 mK increments and
with applied pressures, $P$, as high as $\sim$1.8 kbar.   For all loops, the field sweep rate was $dH/dt= 3.0$ mT/s.  A typical hysteresis
loop (taken at 2.0 K and with no applied pressure) is shown in
the inset of Fig.~\ref{inset}. The magnetization steps
characteristic of resonant tunneling are apparent.   The loop
also displays features near zero field, which are the
paramagnetic response of the fast-relaxing species as well as
of some magnetic impurities in the apparatus' materials.  The
main figure shows a close-up of the $N = 4$ step for this loop. 
For every tunneling step clearly visible in each loop, we empirically
fit the data with a hyperbolic tangent plus a straight line (to
adjust for residual background effects) to determine
$H_N(T,P)$, the field center for step $N$ (the field at which
the tunneling rate is maximum). The main part of
Fig.~\ref{inset} shows an example of such a fit, superimposed
on the data.  For a given $N$ and $T$, the step centers
generally shift toward higher field with increasing pressure.
Figure~\ref{hp2k} shows $H_N$ as a function of pressure for the
$N = 4$ step at 2.0 K.  With the exception of cases where the
step is very small, with a correspondingly small
signal-to-noise ratio, $H_N$ appears to depend linearly on
pressure over the range of pressures studied. The results show
no apparent baric hysteresis as evidenced by the fact that some
of the data at $\sim$0.45 kbar and all of the data at 0 and
$\sim$0.67 kbar were taken after applying the highest
pressure.

\begin{figure}[htb]
\centering
\includegraphics[width=85mm]{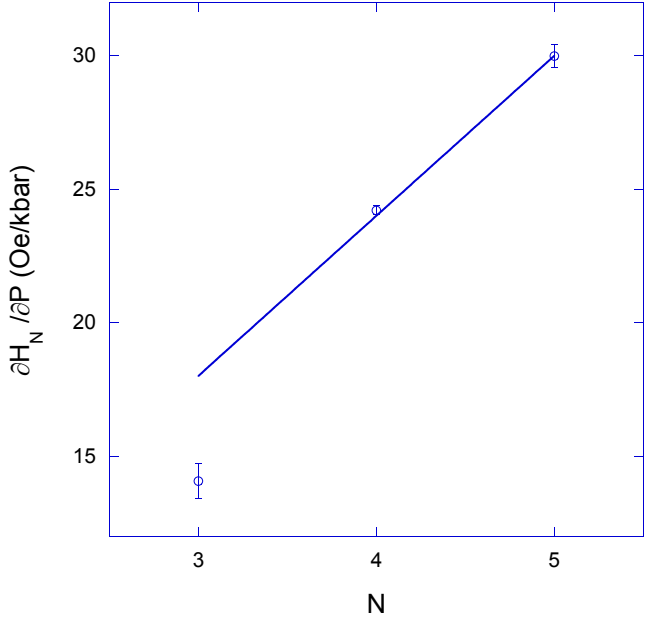} 
\caption{$\partial H_N/\partial P$ as a
function of step number.  The line is the result of a linear fit with fixed
zero intercept. \label{dhdpvsn} }
\end{figure}

The values of $H_N$ depend on both pressure and temperature.
The latter dependence, involving both changes in tunneling levels ($m$ and $m'$ in Eq.~\ref{nfield})~\cite{372, 322} and
the effects of mean dipolar fields~\cite{20}, is significantly
stronger than the effect of pressure.  To determine the effect
of pressure, we 
take $H_N$ for each step
to be a linear function of pressure with a different
zero-pressure intercept for each temperature:
\begin{equation}
\begin{aligned}
H_N (T,P) = H_N (T,0) + \frac{{\partial H_N }}{{\partial P}}P \label{tpfit}
\end{aligned}
\end{equation}
For each step $N$, we simultaneously fit all obtained values of $H_N (T,P)$ (for all values of $T$ and $P$) to Eq.~\ref{tpfit}, to obtain a
value of $\partial H_N/\partial P$ for that step.  Figure~\ref{dhdpvsn} shows
the values of $\partial H_N/\partial P$ as a function of $N$.  (Here and below we have grouped together data for $N$ and $-N$.) The figure also
shows a fit to a line with zero intercept. This result suggests that the
anisotropy parameter $D$ is increasing linearly with applied
pressure, consistent with Eq.~\ref{nfield}, which predicts  $\partial H_N/\partial P=\frac{{N}}{{g\mu _0 \mu _B }}dD/dP$ when $dA/dP=0$.  From the fit, we obtain a
pressure-induced relative change of $D$ of +0.142(1)$\%$/kbar.
This value differs somewhat from those found in experiments
employing hydrostatic pressure~\cite{858, 111}. The difference
may be due to the uniaxial nature of the applied pressure in
our study.

We determined the values of $m$ (and $m'=N-m$) for each step by carefully ascertaining the values of $H_N$ at zero pressure and, after correcting for dipole-field effects~\cite{note2}, employing Eq.~\ref{nfield}.  From this analysis we find that the pair of levels ($m,m'$) between which tunneling occurs are three levels below the top of the classical barrier for all three steps ($N$ = 3,4,5) that we measured~\cite{note}.

As a consistency check of the above interpretation that pressure is inducing a change in $D$, we looked at how the pressure changes the effective barrier height $U_{eff}$.  To do this, we also measured the maximum slope $dM/dH$ for each step and the corresponding value of $M$ at that point in the step.  Under the approximation that the system has a single relaxation rate $\Gamma$, the relaxation is governed by the simple differential equation
\begin{equation}
\frac{dM}{dt}=\frac{dM}{dH} \frac{dH}{dt}= -\Gamma\left(M-M_{eq}\right),\label{diffeq}
\end{equation}

\noindent where the field sweep rate $dH/dt$ is constant for our experiments.  $M_{eq}$ is the equilibrium magnetization and is nearly equal to the saturation magnetization $M_{sat}$ for the values of $N$ and $T$ investigated.  With that approximation, $\Gamma$ can be determined from Eq.~\ref{diffeq} and the measured values of $dM/dH$ and $M$ for each step:

\begin{equation}
\Gamma=-\frac{\frac{dM}{dH}}{\left(M-M_{eq}\right)}\frac{dH}{dt},\label{gamma}
\end{equation}

At the same time, assuming a linear pressure dependence of $U_{eff}$ (i.e. $U_{eff}=U_{eff,0}+U_{eff}^\prime P$), Eq.~\ref{arr} yields

\begin{equation}
\Gamma=\Gamma_0\left(H,T\right) e^{-U_{eff}^\prime P/T},\label{gammaexp}
\end{equation}


\noindent where the prime indicates differentiation with respect to $P$ and $\Gamma_0\left(H,T\right)$ is the relaxation rate at $P=0$. We neglect the small $P$ dependence $\omega_0$ may have.

\begin{figure}[htb]
\centering
\includegraphics[width=75mm]{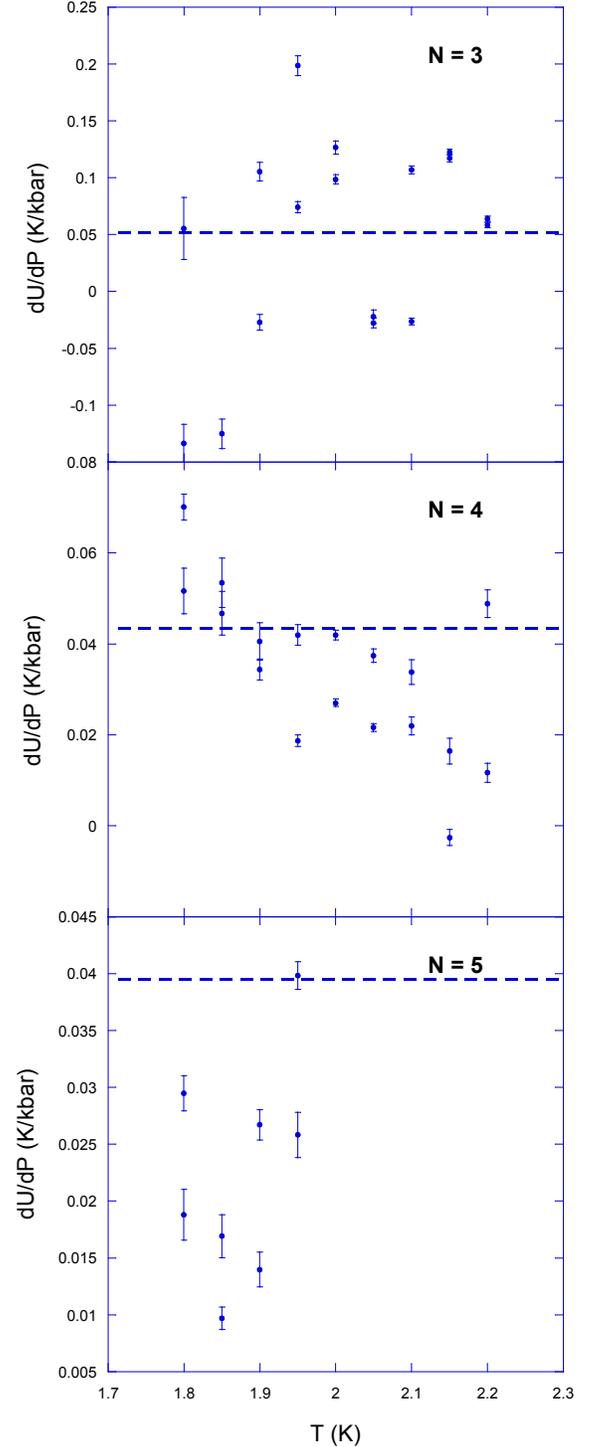} 
\caption{$U_{eff}^\prime$ as a function of $T$ for $N=3,4,5$, as labeled. The horizontal dashed lines are the expected values of $U_{eff}^\prime$ using the value of $dD/dP$ determined from the results in Fig.~\ref{dhdpvsn}.  Note that the three panels have different vertical scales. \label{dudpvst} }
\end{figure}

For each $N$ and $T$, we determine $\Gamma \left(P\right)$ using Eq.~\ref{gamma} and then fit the pressure dependence to Eq.~\ref{gammaexp} to determine $U_{eff}^\prime$.  The results are shown in Fig.~\ref{dudpvst} as a function of $T$.  The horizontal dashed lines are the expected values of $U_{eff}^\prime=dD/dP\left(S^2-m^2\right)$ using the value of $dD/dP$ determined from the results in Fig.~\ref{dhdpvsn} and the value of $m$ we determined from the values of $H_N$ for each $N$~\cite{note}.

The $N=3$ data is very scattered and we can only conclude from it that the value of $U_{eff}^\prime$ agrees on average with the sign and order or magnitude of the prediction.  For $N=4$, the scatter is smaller (note the change in vertical scale) and the agreement with the prediction is better, although there is an unexpected decrease with increasing $T$.  For $N=5$, the data is consistently below the predicted value, but, again, of the same order of magnitude.  The disagreements most likely arise from the assumptions of a single relaxation rate $\Gamma$ (Eq.~\ref{diffeq}) and monodispersed values of $D$ and $A$.  It is well known that Mn$_{12}$Ac has a distribution of relaxation rates due to a distribution in anisotropy parameters as well as tilts of the easy axis away from the crystallographic symmetry axis~\cite{750,317,299,226,122}.  These effects result in a broadening of the tunneling steps that effectively creates a maximum value of $dM/dH$.  Thus, when the steps are large (higher-$T$ data for $N=4$ and all data for $N=5$), the value of $\Gamma$ estimated using Eq.~\ref{gamma} is nearly saturated, suppressing its $P$ dependence.

While it is hard to draw strong conclusions from the results in Fig.~\ref{dudpvst}, it is clear that this data is generally consonant with the interpretation that pressure is inducing a change in $D$ and consistent with the hypothesis that $dA/dP=0$.  If we consider that $A$ is also changing with pressure, it is difficult to devise a situation with significantly better agreement with the data.  For example, allowing $dA/dP=0.5$\%/kbar (reducing $dD/dP$ to 0.116(1)\%/kbar) results in a very modest improvement in the fit in Fig.~\ref{dhdpvsn}, but roughly doubles the predicted values of $U_{eff}^\prime$ in Fig.~\ref{dudpvst}.  If we take an extreme case and fix $dD/dP=0$, then a fit of the data in Fig.~\ref{dhdpvsn} results in $dA/dP=2.73(1)$\%/kbar, which predicts $U_{eff}^\prime$ to be in the range 0.27--0.29 K/kbar, clearly inconsistent with the results in Fig.~\ref{dudpvst}.  We cannot rule out that pressure is inducing small changes in $A$, but the only cases in which changes in $A$ would produce significant effects require that its relative change be larger than the relative change in $D$, a somewhat unlikely scenario.

\begin{figure}[hbt]
\centering
\includegraphics[width=85mm]{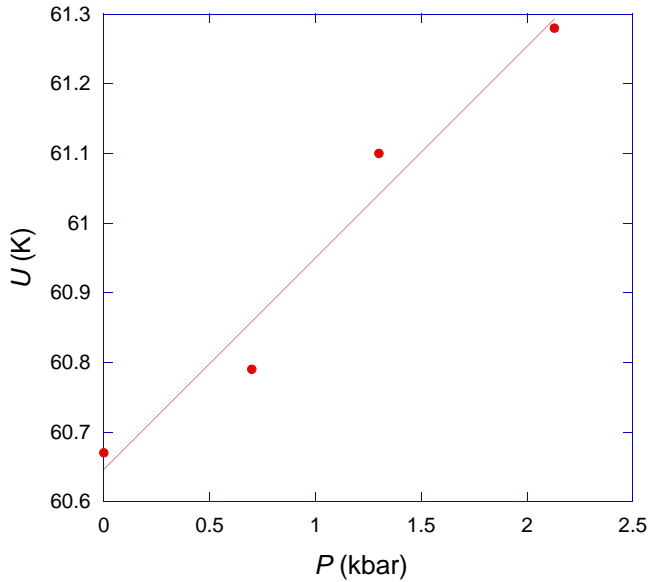} 
\caption{Calculated energy barrier as a function of uniaxial pressure.  Calculations were performed using density-functional theory including spin-orbit coupling, as described in the main text.  The line shows the result of a linear fit to the calculated values.   \label{UvsP} }
\end{figure}

To complement the experimental findings, we performed \emph{ab initio} calculations of the pressure dependence of the energy barrier of Mn$_{12}$. We calculated the electronic structure and magnetic properties of the molecule using density-functional theory (DFT) including spin-orbit coupling (SOC). We used the plane-wave-based DFT code VASP~\cite{861,862} and, for the exchange-correlation potential, we used the Perdew-Burke-Ernzerhopf 
generalized-gradient approximation
~\cite{863}. Projector-augmented-wave 
pseudopotentials were used. The kinetic-energy cutoff was set to 580 eV. For our calculations, we used a unit cell of dimensions 24 x 24 x 24 {\AA}$^3$ containing one Mn$_{12}$ molecule. We performed self-consistent DFT calculations including SOC until the total energy converged to within 0.5 $\mu$eV. 
To simulate the effect of uniaxial pressure, we assumed uniform strain along the z axis of the molecule.  This was implemented by contracting all z coordinates of the molecule by a given factor (up to 1.5\%) while leaving the x and y coordinates unchanged. For each value of strain, we computed the magnetic anisotropy barrier $U$ in zero magnetic field.  
We converted strain to applied pressure using the method discussed in~\cite{Nielson85a,Nielson85b} (tantamount to a Young's modulus of 137.1 kbar). The results of these calculation are shown in Fig.~\ref{UvsP}.

These calculations indicate a qualitatively similar increase in $U$ with pressure as observed experimentally.  Fitting the data in Fig.~\ref{UvsP} to a straight line, yields a slope of $dU/dP$ = 0.30(4) K/kbar, about a factor of four larger than predicted: $dU/dP=S^2 dD/dP =0.0790(5)$ K/kbar.  The discrepancy is not surprising: It indicates that most of the induced strain is intermolecular, which is explicitly not considered in the theoretical model.  In other words, the Young's modulus for the crystal is significantly smaller than that for the molecule itself.

\section{Summary}
We found that by applying uniaxial pressure along the easy axis
of Mn$_{12}$ acetate we were able to increase the magnitude of
the applied field values at which resonant tunneling occurs. We
explain this as an increase in the anisotropy parameter $D$ in the
system's spin Hamiltonian. We determine that pressure causes a
change in $D$ of 0.142(1)$\%$/kbar.  This interpretation is consistent with the pressure-induced changes in the relaxation rate $\Gamma$ as well as \emph{ab initio} calculations of the the effect of pressure on the barrier height $U$.
{\indent}

While in this experiment, the pressure was applied along the
molecule's symmetry axis, one could also apply pressure along a
perpendicular direction, which could affect the tunneling rate
by breaking the symmetry of the molecule.  In fact, a
theoretical calculation predicts that pressure applied along
the hard axis of a four-fold symmetric molecule could modulate
the tunneling rate via a geometric-phase effect~\cite{30}.
Future experiments may attempt to observe this effect.

\acknowledgments We would like to acknowledge the valuable
programming and analysis contributions of D. Wadden, as well
the contributions of C. Mochrie, J. Rasowsky and D. K. Kim for
their work on previous versions of this experiment. We are also
indebted R. Cann and D. Krause for their essential aid in the
design and construction of the apparatus, and J. Kubasek for his help in producing Fig.~\ref{apparatus}.  Support for this
work was provided by the National Science Foundation under
grant nos.~DMR-0449516 and DMR-1006519, and by the Amherst College Dean of Faculty.

\bibliographystyle{eplbib}
\bibliography{jha}

\begin{thebibliography}{10}
\bibitem{note2}
Explicitly, we used $H_N^{corrected} = H_N^{measured} + \alpha
  M\left(H_N\right)/M_{sat}$, where 
  $M_{sat}$ is the saturation magnetization and $\alpha$ = 515 Oe~\cite{20}.
\bibitem{note}
To be precise, we find that for $N=3,4,5$, respectively, the values
  of the pairs of levels involved in tunneling $(m,m')$ are (-4,1), (-5,1) and
  (-5,0). The pair of tunneling levels for Mn$_{12}$ can depend on
  temperature~\cite{372,322}; however, over the narrow temperature range in our
  study, we do not expect this effect to be very significant. In any case, a
  change of the $N$ = 4 pair, for example, from (5,-1) to, say, (4,0) would not
  have much effect on our findings. Such a change has no effect on the results
  for $\partial H_N/\partial P$ and has
  very modest effects on the analysis of $dU/dP$, raising the expected value in
  Fig.~\ref{dudpvst} (middle panel) from 0.043 to 0.047 K/kbar.
\end{thebibliography}
\end{document}